\documentclass[cits]{PoS}
\newcommand{\preprint}{\newline
  \begin{picture}(0,0)
  \put(0,67){\rm\footnotesize DESY 09-198, HU-EP-09/56, IFT-UAM/CSIC-09-55, LTH853, ROME1/1467/2009, RM3-TH/09-20, SFB/CPP-09-109}
  \end{picture}}

\newcommand{\fm}{\, {\rm fm}}

\newcommand{\mev}{\, {\rm MeV}}

\newcommand{\be}{\begin{equation}}
\newcommand{\ee}{\end{equation}}
\newcommand{\bea}{\begin{eqnarray}}
\newcommand{\eea}{\end{eqnarray}}
\newcommand{\nn}{\nonumber}
\newcommand{\bi}{\begin{itemize}}
\newcommand{\ei}{\end{itemize}}

\boldmath
\title{$f_B$ and $f_{B_s}$ with maximally twisted Wilson fermions\preprint}
\unboldmath

\ShortTitle{$f_B$ and $f_{B_s}$ with tmQCD}

\author{\Large{ETM Collaboration}}
\author{B.~Blossier\\
Laboratoire de Physique Th\'eorique (B\^at.~210), Universit\'e de
Paris XI,\\ Centre d'Orsay, 91405 Orsay-Cedex, France}
\author{P.~Dimopoulos, R.~Frezzotti, G.~C.~Rossi\\
Dip. di Fisica, Universit{\`a} di Roma Tor Vergata and INFN, Sez.
di Roma Tor Vergata,\\ Via della Ricerca Scientifica, I-00133 Roma, Italy}
\author{G.~Herdoiza, K.~Jansen\\
NIC, DESY, Platanenallee 6, D-15738 Zeuthen, Germany}
\author{V.~Lubicz, C.~Tarantino\\
Dip. di Fisica, Universit{\`a} di Roma Tre, Via della Vasca Navale
84, I-00146 Roma, Italy\\
INFN, Sez. di Roma
Tre, Via della Vasca Navale 84, I-00146 Roma, Italy}
\author{G.~Martinelli, F.~Sanfilippo\\
INFN, Sezione di Roma, I-00185 Roma, Italy\\
        Dipartimento di Fisica, Universit\`a di Roma ``La
        Sapienza'', I-00185 Roma, Italy}
\author{C.~Michael, A.~Shindler\footnote{Current address: Instituto de F\'{\i}sica Te\'orica UAM/CSIC
Universidad Aut\'onoma de Madrid, Cantoblanco E-28049 Madrid, Spain}\\
Theoretical Physics Division, Dept. of Mathematical Sciences,
\\University of Liverpool, Liverpool L69 7ZL, UK}
\author{S.~Simula\\
INFN, Sez. di Roma
Tre, Via della Vasca Navale 84, I-00146 Roma, Italy}
\author{C.~Urbach, M.~Wagner\\
Humboldt-Universit\"at zu Berlin, Institut f\"ur Physik, Newtonstra{\ss}e
15, D-12489 Berlin, Germany}

\abstract{We present a lattice QCD calculation of the heavy-light decay constants
$f_B$ and $f_{B_s}$ performed with $N_f=2$ maximally twisted Wilson
fermions, at four values of the lattice spacing. The decay constants have been also computed in the static
limit and the results are used to interpolate the
observables between the charm and the infinite-mass sectors, thus
obtaining the value of the decay constants at the physical $b$ quark
mass. Our preliminary results are $f_B=191(14) \mev$, $f_{B_s}=243(14) \mev$, $f_{B_s}/f_B=1.27(5)$.
They are in good agreement with those obtained with a novel
approach, recently proposed by our Collaboration (ETMC), based on the
use of suitable ratios having an exactly known static limit.}

\FullConference{The XXVII International Symposium on Lattice Field
  Theory - LAT2009\\ July 26-31 2009\\ Peking University, Beijing,
  China}

\begin{document}

\section{Introduction}
 The study of B-physics plays a fundamental role within flavour physics
both in accurately testing the Standard Model and in the search of
New Physics effects. To this aim it is crucial to have theoretical
uncertainties under control, in particular those of the hadronic
parameters computed on the lattice.

With the available computer power it is not possible to simulate quark
masses in the range of the physical $b$ mass keeping, at the same time,
finite volume and discretisation effects under control. In order to
circumvent these problems, many different methods have been proposed so
far (see ref.~\cite{Aubin:2009yh} for an up to date collection of
results). 

The approach that we have adopted and that we discuss below consists in
using lattice QCD data with the heavy quark mass ranging from the charm region up to $\sim 4/5$ of the physical $b$ quark mass,
together with the information coming from the static limit point. In order to deal with the simulated light
quark mass and finite lattice spacing, a careful extrapolation to the
chiral and continuum limits has been performed. An
alternative method, based on the introduction of suitable ratios having
an exactly known static limit, has been recently proposed and
investigated by our Collaboration (ETMC)~\cite{Blossier:2009hg}.

In section \ref{sec:static} we describe the computation of the decay
constants in the static limit; in section \ref{sec:standard} we present
the interpolation between the charm and
infinite-mass sectors and compare the results with those obtained in ref.~\cite{Blossier:2009hg}.

\section{Heavy-light decay constant in the static limit of HQET} \label{sec:static}

We have combined a light doublet of twisted-mass fermions ($\psi^T =
(u,d)$) defined at maximal twist with a static quark described by the
HYP2 action~\cite{HasenfratzHP} to improve the signal-to-noise  ratio~\cite{DellaMorteYC}:
 \begin{equation}
S^{\rm stat}= a^4 \sum_x \bar{\psi}_{\rm h}(x) \nabla^*_0 \psi_{\rm h}(x), \quad
\nabla^*_0 \psi_{\rm h}(x) = \frac{1}{a} \left[\psi_{\rm h}(x) - U^\dag_{\rm HYP2}
(x-a\hat{0})\psi_{\rm h}(x-a\hat{0})\right].
 \end{equation}
 In order to extract the decay constant using maximally twisted lattice
QCD,  we need to evaluate the matrix element of the static-light local
current. At maximal twist the pseudoscalar
current $\left({\mathcal P}^{\rm stat}\right)_{R}$  in the  physical basis, in terms of  the
twisted basis used in the numerical simulations (light quark fields
$\chi^T=(\chi_u, \chi_d)$), is given by
 \be
\left({\mathcal P}^{\rm stat}\right)_{R}(x) = \left(\bar{\psi}_{\rm h}(x) \gamma_{5} u(x)\right)_R = \frac{1}{\sqrt2}\left(Z_P^{\rm stat} P(x) + i\,Z_S^{\rm stat} S(x)\right)
 \ee
 where $P=\bar{\psi}_{\rm h} \gamma_{5} \chi_u$ and $S = \bar{\psi}_{\rm
h}\chi_u$ are the pseudoscalar and scalar densities which renormalise
with the  $Z_P^{\rm stat}$ and $Z_S^{\rm stat}$ appropriate to the
static-light framework.

We define $c_1 = i\,\langle 0 | \bar{\psi}_{\rm h}\chi_u | B \rangle$
and  $c_5 = \langle 0 | \bar{\psi}_{\rm h}\gamma_5\chi_u | B \rangle$
where $|B\rangle $  is the lattice ground state. At maximal twist,
the amplitude we  need to compute is $\Phi = f_B \sqrt{M_B}=
\left(Z_S^{\rm stat} c_1 + Z_P^{\rm stat} c_5 \right)$. The (bare)
matrix elements $c_1$ and $c_5$ have been measured from an analysis
following the static HQET spectrum study with twisted-mass
fermions~\cite{JansenSI}. The ETMC ensembles $B_{1,2,3,4}$ and 
$C_{1,2}$~\cite{BoucaudXU,scaling} have so far been considered (i.e. two
lattice spacings). Here we concentrate on the lightest heavy-light meson
state, the pseudoscalar  meson which we call here the $B$ meson (or
$B_s$ with a strange valence quark).  We take the value of $m_q$ for the
strange quark from the ETMC studies of  the strange-light
mesons~\cite{Blossier:2007vv,Blossier:2009bx} which used the same gauge
configurations as used  here, namely $a m_s = 0.022$ at $\beta = 3.9$
and $0.017$ at $\beta = 4.05$.  We measure the correlation of operators
at source and sink with a large choice  of operators: local and smeared;
parity conserving and non-conserving. We then  make a simultaneous fit
to a sub-matrix (typically $6 \times 6$ ) in a given Euclidean time $t$
interval. We chose this $t$-interval to have similar physical extent at
different lattice spacings. We find  that the non-local operators have
weaker coupling to excited states, as expected.  Such non-local
operators can give a good determination of the energy levels but  to
extract the required matrix element (related to $f_B$ ) we need to
include local  operators in the fit. 
 At $\beta = 3.9$ we use a $4$ state fit with $t/a$ range $4-10$ but
with the correlations  that have local operators (at sink and/or source)
we restrict to $t/a$ range to $6-10$.  This choice gives acceptable
values of $\chi^2$ using correlated fits. We then make  uncorrelated
fits to determine the required energies and matrix elements with 
statistical errors determined by bootstrap. At $\beta = 4.05$ the
appropriate $t/a$ range  is found to be $5-12$ for smeared correlators and
$7-12$ for local ones. We have  checked by making many different fits
that the fit parameters are stable, within  the statistical error
assigned.  For the correlations of $B_s$ mesons, we make similar fits
but find that  the minimum $t/a$ value has to be increased by 1 unit to
preserve an acceptable  (correlated) $\chi^2$ . 

\begin{figure}[tb]	
  \includegraphics[width=0.41\textwidth,angle=270]{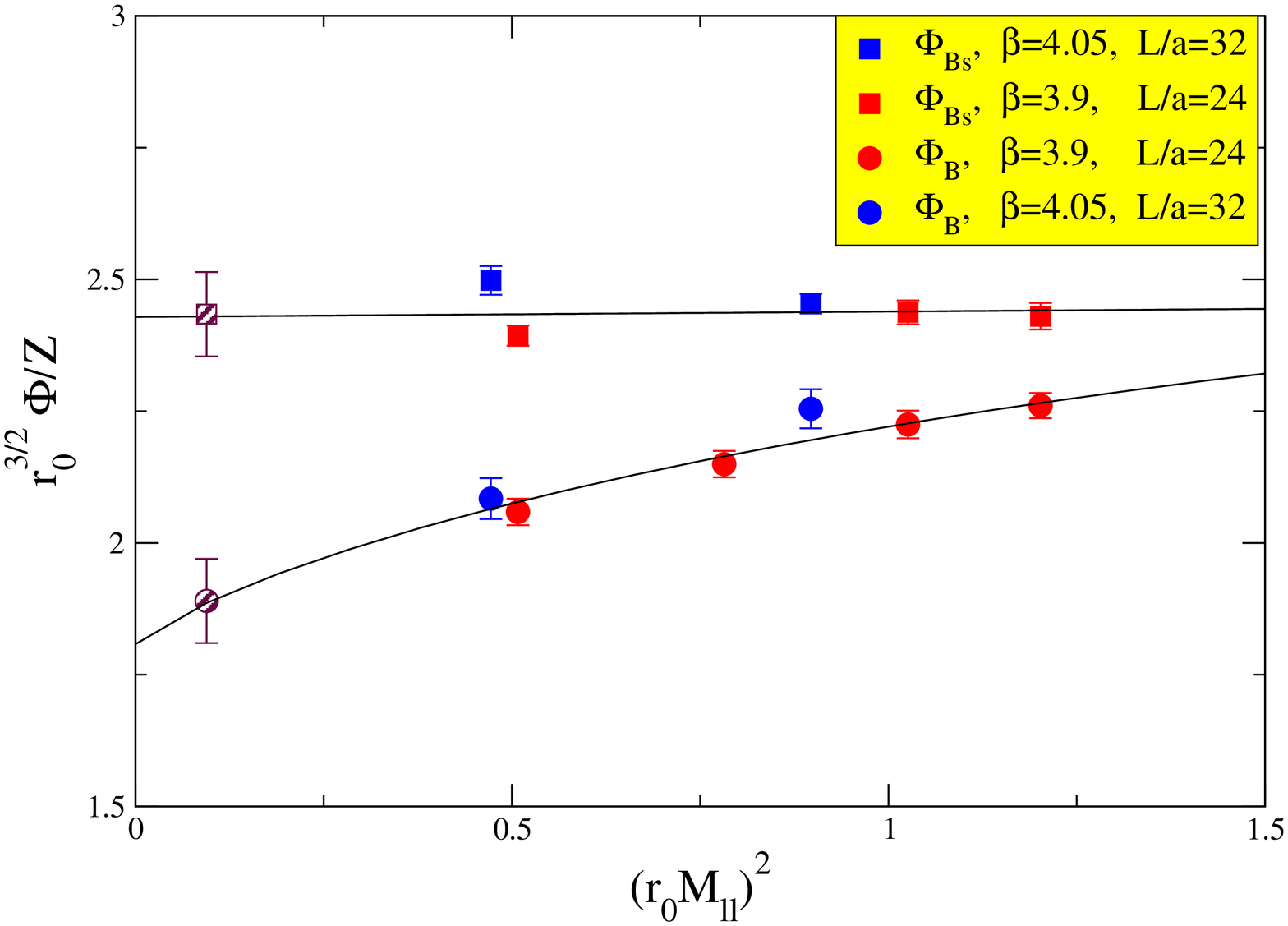}
\includegraphics[width=0.41\textwidth,angle=270]{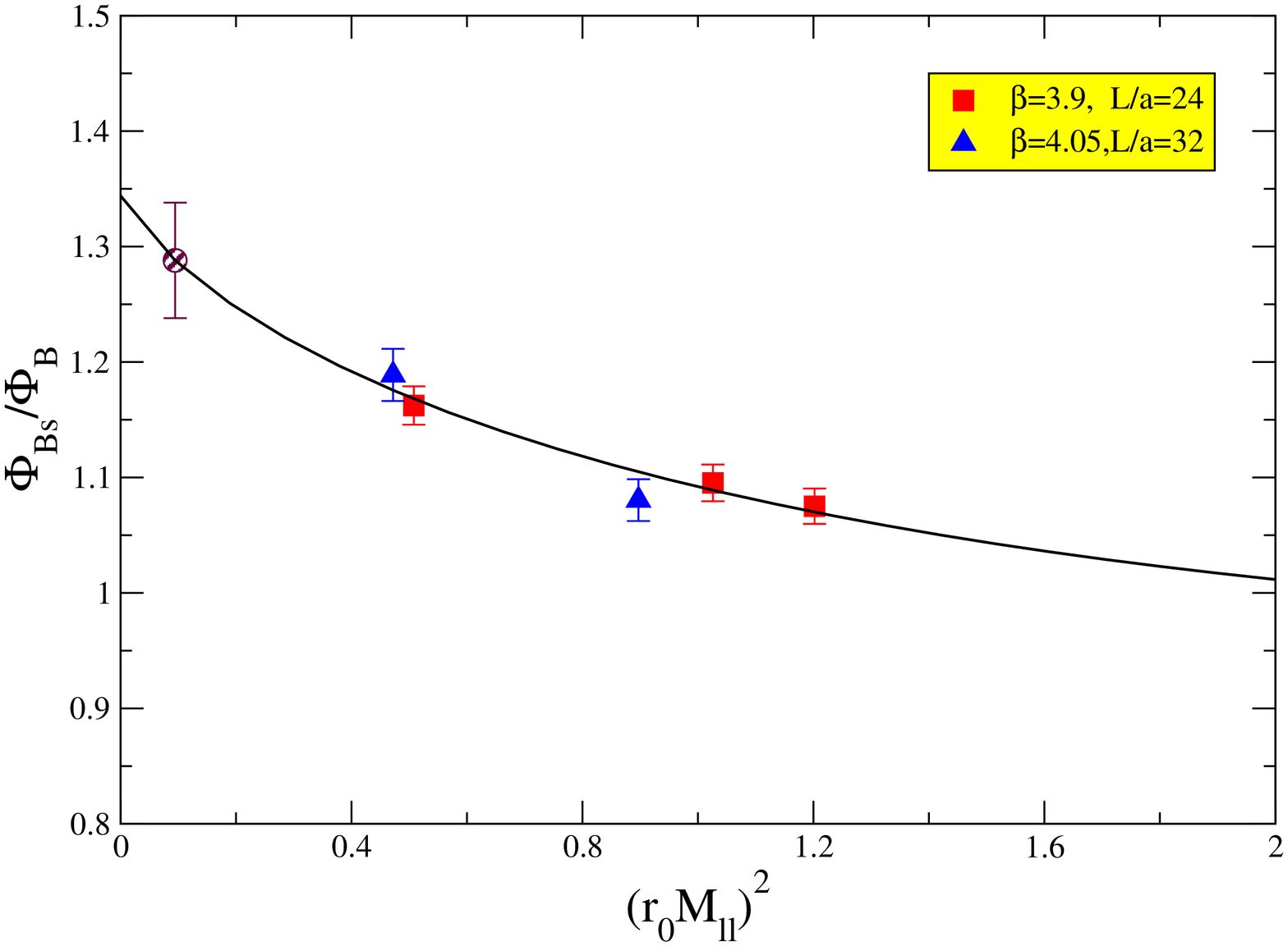}
   \caption{Left plot: unrenormalised heavy-light decay constant
 combination $r_0^{3/2}  \Phi/Z (4.05)$ (with $Z \equiv \left(Z_P^{\rm stat} + Z_S^{\rm stat}\right)/2$) 
versus the squared mass of the pion built of the light sea quarks. 
The circles represent the $B$ meson case, where the valence light quark is equal to the sea quark. 
The squares represent the $B_s$ meson case, where the light valence
quark is the strange quark. The data at  $\beta = 3.9$ (red symbols) have been
multiplied by the appropriate factor to match the same scale for the
data  at $\beta=4.05$. The curves represent the NLO HMChPT theory expressions. Right plot: the ratio $\frac{\Phi_{B_s}}{\Phi_B}$
versus the squared mass of the pion built of the light sea quarks. The
curve represents the NLO heavy quark chiral perturbation theory.
}
  \label{fig:chiralfit}
 \end{figure}

Then one computes $Z_{P}^{\rm stat}$ and $Z_{S}^{\rm stat}$ in order to
get the matrix element renormalised in  HQET at a specific scale $\mu$.
We have chosen to renormalise it in the $\overline{\rm MS}$ scheme at 
$\mu=1/a$ and for this preliminary account of our work the
renormalisation is  done perturbatively at 1 loop order. $\overline{\rm
MS}$ is a  continuum-like scheme defined within dimensional regularisation, while
the regulator of our bare  quantities is the inverse  lattice spacing.
So one needs a matching between both regularisations. It can be written as
 \begin{eqnarray}\nonumber
\langle O(p,\mu) \rangle^{\mbox{DR},\overline{\mbox{MS}}} &=& \left[1 - \frac{\alpha_{s}}{4\pi} 
\left(-\gamma_0 \ln a^2\mu^2 + C^O\right) \right]
\langle O(p,a) \rangle^{\rm lat}\\
\label{matching}
&\equiv&Z_O(a\mu) \langle O(p,a) \rangle^{\rm lat}\,,
 \end{eqnarray}
where the renormalisation scheme and scale of the coupling constant $\alpha_s$ is not specified at this level of perturbation theory.
Expressions of
$C^{P(S)}$ are complicated and not illuminating,  essentially
due to the HYP-smeared static action and the improved part of  the gluon
propagator~\cite{Horsley:2004mx}. Thus we have simply collected  the
numerical values of $Z_{P}^{\rm stat}$ and $Z_{S}^{\rm stat}$ in Table
\ref{Zstats} for a  boosted coupling $g_P^2=g_0^2/\langle U_P \rangle$
(where $g_0^2 = 6 /\beta$ and  $\langle U_P \rangle$ is the average
plaquette value).
 \begin{table}[b]
\begin{center}
\begin{tabular}{|c|c|c|}
\hline
$\beta$&$Z_{P}^{\rm stat}$&$Z_{S}^{\rm stat}$\\
\hline
3.9&0.849&0.933\\
4.05&0.859&0.938\\
\hline
\end{tabular}
\end{center}
 \caption{\label{Zstats} First order perturbation theory renormalisation
factors of the pseudoscalar and scalar  static-light
dimension 3 operators in the $\overline{MS}$ scheme at the  scale $\mu=1/a$.}
 \end{table}
It turns out that the systematic error introduced by a poor
determination of the ratio $z_r = Z_S^{\rm stat}/Z_P^{\rm stat}$
is minimal, especially on the ratio of the $B$ and $B_s$ decay constants.
We thus present in fig.~\ref{fig:chiralfit} the bare matrix element, which depends on the ratio $z_r$ only.

 Once the matrix element $\Phi^{\overline{MS}}(\mu=1/a)$ has been
renormalised in the $\overline{MS}$ scheme at the scale $\mu=1/a$ a NLO
running of perturbation theory~\cite{ChetyrkinVI} has been applied  to
evolve it to a scale $\mu = M_B^{\rm exp}$. This is what is needed to
perform a fit together with the relativistic data matched to HQET at the
same  scale (see next section).

The extrapolation of $\Phi_B$ down to the physical pion has been
performed with Heavy Meson Chiral Perturbation Theory (HMChPT) at
NLO by using the formula~\cite{GoityTP, GrinsteinQT, Sharpe:1995qp}
\bea
\frac{\Phi_B}{\Phi_0} &=& 1 - \frac{3(1+3\hat{g}^2)}{4} \frac{M^2_{ll}}{(4\pi f)^2} 
\log \left(\frac{M^2_{ll}}{(4\pi f)^2}\right)+\alpha_1\,M^2_{ll}\,,\nn\\
\frac{\Phi_{B_s}}{\Phi_{0s}} &=& 1 +\alpha_{1s}\,M^2_{ll}\,,
\label{eq:HMChPTstatic}
\eea
 where $M_{ll}$ denotes the simulated pion masses, $f$ stands for the light decay constant in the chiral limit, while $\Phi_{0(s)}$ and $\alpha_{1(s)}$ are free fit parameters. The $\hat{g}^2$
coupling has been fixed to  $0.2$~\cite{OhkiPY, BecirevicYB}, and we have
checked that a change of $50\%$ in the value of $\hat{g}^2$  results in
a shift in $\Phi_B$ which is well below  the statistical error.
 The chiral extrapolation of $\Phi_B$, $\Phi_{B_s}$ and the ratio
$\Phi_{B_s}/\Phi_B$ is shown in  fig.~\ref{fig:chiralfit}. This figure
also illustrates that we find consistent results  at our two available
lattice spacings within the relatively large errors. We do not have
enough data to include explicit discretisation error terms in the fit
formula. However it seems that cut-off effects are quite small. This is
more evident for the ratio $\frac{\Phi_{B_s}}{\Phi_B}$ which is
consistent with having no cutoff effects (see right plot of
fig.~\ref{fig:chiralfit}).

\boldmath
\section{Relativistic results and interpolation to the physical $b$ quark mass} \label{sec:standard}
\begin{figure}[tb]	
  \center{\includegraphics[width=0.46\textwidth,angle=270]{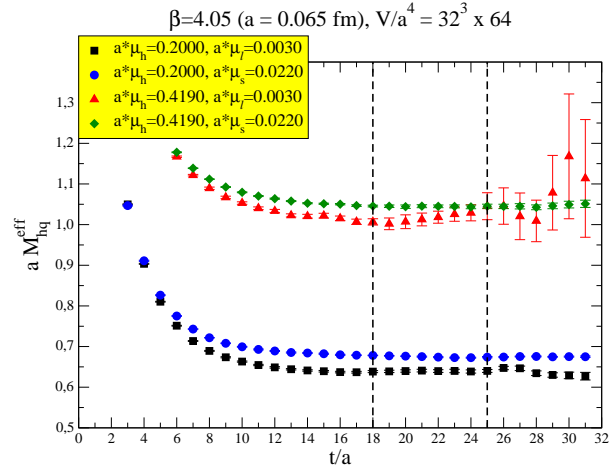}}
  \caption{Effective masses at $\beta=4.05$ for two heavy-light ($hl$) and two
    heavy-strange ($hs$) quark mass combinations. The two heavy quark masses correspond approximately to the physical charm quark mass and to $\sim 2/3$ of the value of the physical $b$ quark mass.}
  \label{fig:plateaux}
\end{figure}

\unboldmath
We perform an interpolation of the heavy-light
($hl$) decay constants from the charm region up to the bottom mass, by
including data in the static limit calculated in the HQET as explained
in the previous section. The lattice QCD data used in this analysis are at
four values of the lattice spacing $a \approx 0.100, 0.085, 0.065, 0.050
\fm$ (corresponding to $\beta=3.8, 3.9, 4.05, 4.2$), that is we have used the configuration ensembles denoted in~\cite{BoucaudXU,scaling} as $A_{2,3}$, $B_{1,2,3,4,6,7}$, $C_{1,2,3}$ and $D_2$, respectively.
We have simulated for each ensemble 16 heavy quark masses in the range $m_c^{phys} \lesssim m_h \lesssim 0.8
m_b^{phys}$. Quark propagators with different valence masses are obtained using the so called multiple mass solver method~\cite{Jansen:2005kk}. In fig.~\ref{fig:plateaux} we show for illustrative
purpose the effective masses at $\beta=4.05$ and for few quark mass
combinations.

The analysis is performed by studying the dependence of the decay
constants, more precisely of the quantity $\Phi_{hq}=f_{hq} \sqrt{M_{hq}}$, as a function of the meson masses, as in our recent analysis of the $f_D$ and $f_{D_s}$ decay constants~\cite{Blossier:2009bx}.

In order to make use of the HQET scaling low we introduce for each simulated $hq$ meson mass $M_{hq}$ the HQET quantity that is finite in the static limit~\cite{ChetyrkinVI}:
\bea
\Phi_{hq}&=&\Biggl(\frac{\alpha^{\overline{\mbox{MS}}}(M_{hq})}{\alpha^{\overline{\mbox{MS}}}(M_B^{\rm exp})}\Biggr)^{-\gamma_0/(2\beta_0)}\,\cdot
\Biggl[1-\Biggl(\frac{439}{1089}-\frac{28 \pi^2}{297} \Biggr)\, \frac{\alpha^{\overline{\mbox{MS}}}(M_{hq})-\alpha^{\overline{\mbox{MS}}}(M_B^{\rm exp})}{4\pi}\Biggr]\cdot\nn\\
&&\cdot\Biggl[1+\frac{8}{3}\frac{\alpha^{\overline{\mbox{MS}}}(M_{hq})}{4\pi}\Biggr]\,\cdot(\Phi_{hq})_{QCD}\,,
\eea
which has been obtained through the NLO matching from QCD to HQET and evolving at NLO to the renormalisation scale given by the experimental value of the $B$ meson mass.
For $\Phi_{hq}$ ($q=l,s$) we first study the dependence on the light/strange
quark mass at fixed heavy mass through the following functional forms 
\bea 
\Phi_{hl} &=&
A(a, m_h) \cdot \left(1-\frac{3\,(1+3 {\hat g}^2)}{4}\cdot \frac{M^2_{ll}}{(4\pi f)^2} \cdot \log
\left(\frac{M^2_{ll}}{(4\pi f)^2}\right)+B\cdot M^2_{ll}\right)\,,\nn\\
\Phi_{hs} &=& A^\prime(a, m_h) \cdot \left(1+ B^\prime \cdot
M_{ll}^2 +C^\prime(a) \cdot M_{ss}^2\right)\,.
\label{eq:phi}
\eea 
We note that the fit forms above follow from the HMChPT formulae~\cite{GoityTP, GrinsteinQT, Sharpe:1995qp}, which we have already used in the static sector (see eq.~(\ref{eq:HMChPTstatic})).
A dependence of the coefficients $A, A^\prime, C^\prime$ on the lattice spacings is allowed, in order to account for discretisation effects.
The extrapolation/interpolation
to the physical light/strange quark mass is performed by replacing in
eq.~(\ref{eq:phi}) $M^2_{ll} = (M_\pi^{exp})^2$, $M^2_{ss} =
2\,(M_K^{exp})^2-(M_\pi^{exp})^2$.  This first step provides the values
of the decay constants at the physical light/strange quark mass for
every simulated lattice spacing and heavy quark mass, or equivalently
the quantities $\Phi_{hq^{phys}}$.

The second step consists in studying the dependence of
$\Phi_{hq^{phys}}$, included the available static points, on the heavy
quark mass and on the lattice spacing, in order to interpolate to the
$b$ quark mass and to extrapolate to the continuum limit.  Several
functional forms with different $\mathcal{O}(a^2)$ and
$\mathcal{O}(a^4)$ discretisation terms have been tried, which can be written in a compact way as
\be
\Phi_{hl^{phys}(hs^{phys})} = \sum_{n,k} P_{nk}\, a^{2n}\,M_{hq}^{2n-k}\,,\qquad (n=0,1,2;\,\,k=0,1,2)\,,
\label{eq:phichiral}
\ee
 where $M_{hq}$ is a reference meson mass with the same simulated heavy
quark mass as in the fitted quantity $\Phi$ and the light quark mass is
fixed to a similar value for all data.
We have performed correlated fits by assuming the static results uncorrelated with the relativistic data.

The results for the decay constants $f_B$ and $f_{B_s}$ are finally
obtained by replacing in eq.~(\ref{eq:phichiral}) $M_{hq}=M_{hs}=M_{B_s}^{exp}$, setting the lattice spacing equal to zero and performing the matching from HQET back to QCD at NLO.

\begin{figure}[tb]	
\includegraphics[width=0.41\textwidth,angle=270]{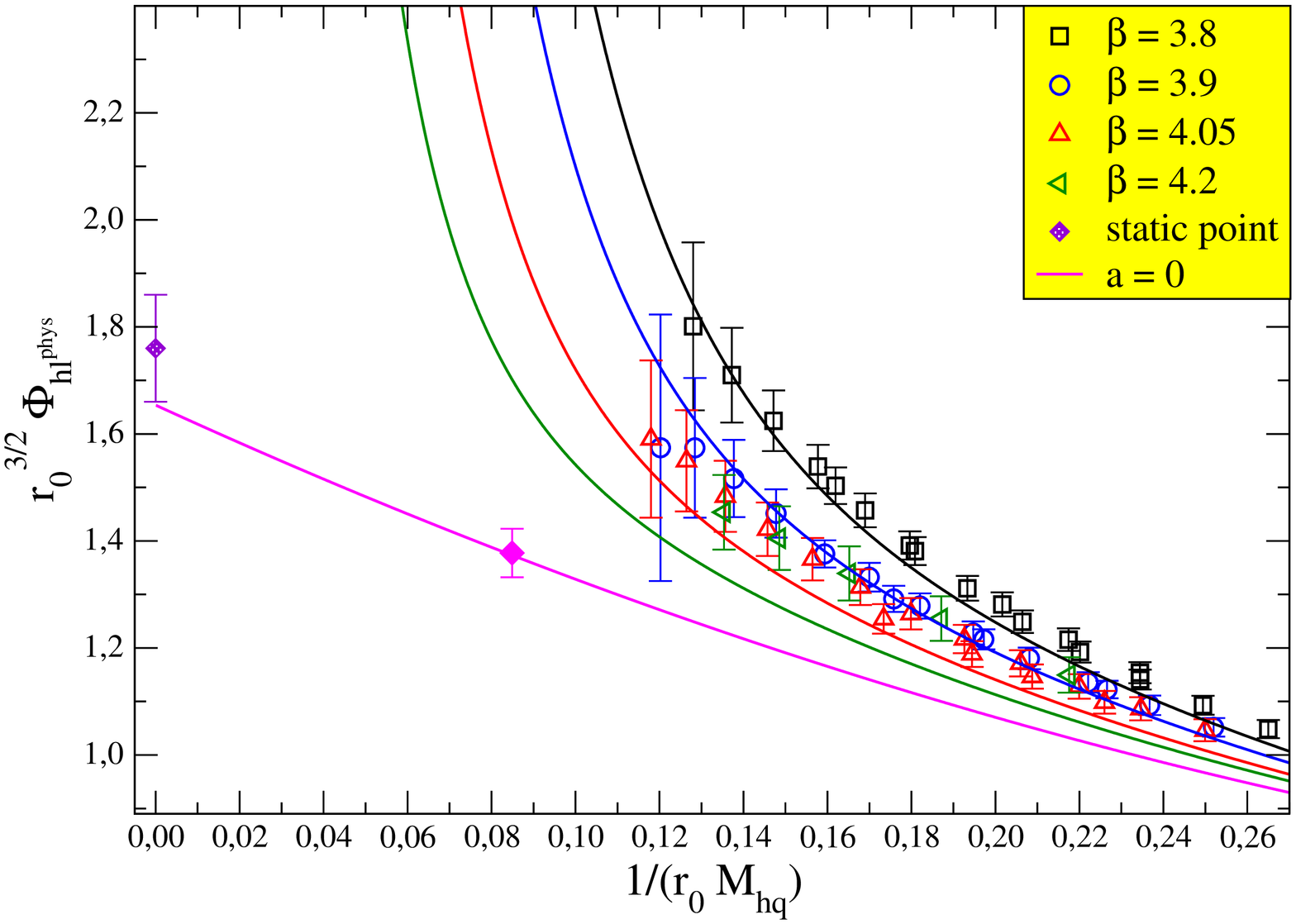}
\includegraphics[width=0.41\textwidth,angle=270]{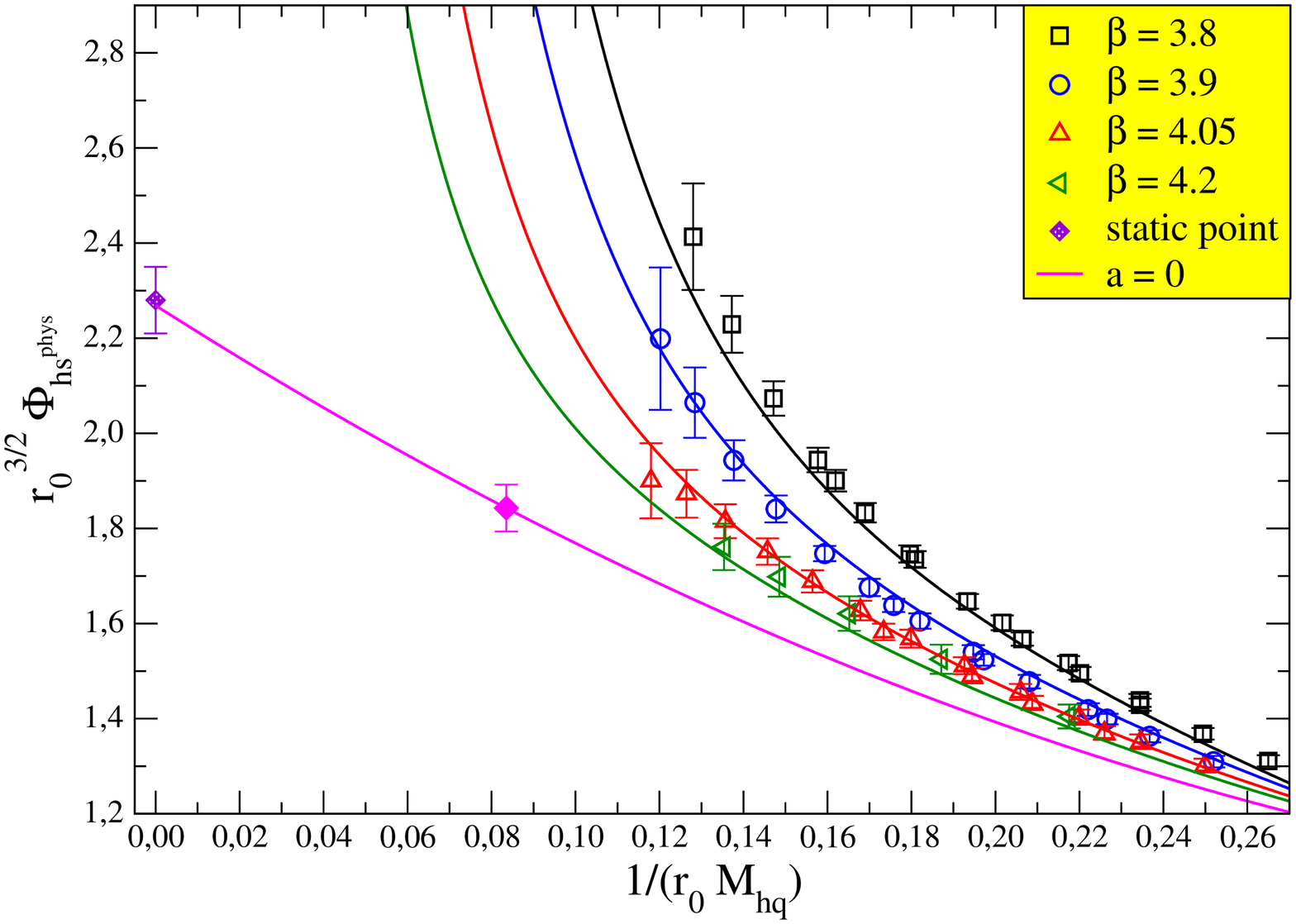}
\caption{Interpolation to the $b$ quark mass and continuum extrapolation
  of $\Phi_{hl^{phys}}$ (left) and $\Phi_{hs^{phys}}$ (right).}
\label{fig:heavy}
\end{figure}

The dependence of the decay constants on the
$hq$ meson mass is shown in fig.~\ref{fig:heavy} where, for illustrative purpose, we also show curves corresponding to one of the various fits.
The discretisation terms included in the shown fits are of $\mathcal{O}(a^2\,M_{hq})$, $\mathcal{O}(a^2\,M_{hq}^2)$ and $\mathcal{O}(a^4\,M_{hq}^4)$ for both $\Phi_{hl^{phys}}$ and $\Phi_{hs^{phys}}$.
We observe that with our data it is not
possible to determine the coefficients of more than three discretisation
terms for each fit and that, in some cases, only two out of three parameters turn out to be different from zero.
About twenty of these fits have a chi square per degree of freedom of order one or smaller and are considered in deriving our final result for $f_B$ and $f_{B_s}$.
The spread among these fits is included in the systematic uncertainty.

Our preliminary results for $f_B$, $f_{B_s}$ and the ratio read\footnote{The results given in the present proceedings are based on a larger statistical sample  w.r.t to the values presented at the Conference and cited in ref.~\cite{Aubin:2009yh}.}
\bea
f_B &=& 191(6)(12)(3) \mev = 191(14) \mev\,,\nn\\
f_{B_s} &=& 243(6)(12)(3)\mev = 243(14) \mev\,,\nn\\
f_{B_s}/f_B &=& 1.27(3)(4) = 1.27(5)\,,
\label{eq:standard}
\eea
where: i) the first error is of statistical plus fitting origin,
ii) the second error, estimated through the spread of the results
obtained with functional forms containing different discretisation
terms, represents the residual uncertainty due to the continuum limit
and to the $b$ mass interpolation, iii) the third error takes into
account the effect of the systematic uncertainty on the static
point.

We conclude by comparing the results in eq.~(\ref{eq:standard}) with those obtained in ref.~\cite{Blossier:2009hg} using suitable ratios having an exactly known static limit.
The latter values read
\bea
f_B &=& 194(16) \mev,\nn\\
f_{B_s} &=& 235(11) \mev\,,
\label{eq:ratio}
\eea
where the uncertainty is the sum in quadrature of the statistical and systematic errors.
The two sets of results are in very good agreement, thus providing further confidence on their robustness. 
We note that the results in eq.~(\ref{eq:ratio}) are obtained from a subset of the data analysed in the present study. 
The inclusion of the full set of data is in program for a forthcoming publication.

\end{document}